\documentclass[11pt,fleqn,oneside,openany]{book} 
\usepackage{amsmath}
\usepackage{amssymb}
\usepackage[top=3cm,bottom=3cm,left=0.8in,right=0.8in,headsep=10pt]{geometry} % Page margins
\usepackage{graphicx} % Required for including pictures

\usepackage{verbatim}
\usepackage{enumitem} % Customize lists
\setlist{nolistsep} % Reduce spacing between bullet points and numbered lists
\usepackage{booktabs} % Required for nicer horizontal rules in tables
\usepackage{xcolor} % Required for specifying colors by name
\usepackage{listings}
\usepackage{color}

\usepackage{anyfontsize}
\usepackage{avant} % Use the Avantgarde font for headings
\usepackage{mathptmx} % Use the Adobe Times Roman as the default text font together with math symbols from the Sym­bol, Chancery and Com­puter Modern fonts
\usepackage{microtype} % Slightly tweak font spacing for aesthetics

\usepackage{calc} % For simpler calculation - used for spacing the index letter headings correctly
\usepackage{makeidx} % Required to make an index
\makeindex % Tells LaTeX to create the files required for indexing

\usepackage{natbib}

\setlist{nolistsep} % Reduce spacing between bullet points and numbered lists

{
     \begin{description}\setlength{\itemsep}{0.15\baselineskip}}
{\end{description}}
 
\def\tasktitle#1{\item{}}

{\begin{enumerate}[label=#1-\arabic{*}.,ref=\thesubsection:#1-\arabic{*},font=\bf]}
{\end{enumerate}}

\usepackage{booktabs}
\usepackage{pdflscape}
\usepackage{float}
\usepackage[colorlinks=true, allcolors=blue]{hyperref}
\usepackage[document]{ragged2e}
\begin{document}

\def\refj@jnl#1{{\rm#1}}

\def\aj{\refj@jnl{AJ}}                   % Astronomical Journal
\def\actaa{\refj@jnl{Acta Astron.}}      % Acta Astronomica
\def\araa{\refj@jnl{ARA\&A}}             % Annual Review of Astron and Astrophys
\def\apj{\refj@jnl{ApJ}}                 % Astrophysical Journal
\def\apjl{\refj@jnl{ApJ}}                % Astrophysical Journal, Letters
\def\apjs{\refj@jnl{ApJS}}               % Astrophysical Journal, Supplement
\def\ao{\refj@jnl{Appl.~Opt.}}           % Applied Optics
\def\apss{\refj@jnl{Ap\&SS}}             % Astrophysics and Space Science
\def\aap{\refj@jnl{A\&A}}                % Astronomy and Astrophysics
\def\aapr{\refj@jnl{A\&A~Rev.}}          % Astronomy and Astrophysics Reviews
\def\aaps{\refj@jnl{A\&AS}}              % Astronomy and Astrophysics, Supplement
\def\azh{\refj@jnl{AZh}}                 % Astronomicheskii Zhurnal
\def\baas{\refj@jnl{BAAS}}               % Bulletin of the AAS
\def\bac{\refj@jnl{Bull. astr. Inst. Czechosl.}}
                % Bulletin of the Astronomical Institutes of Czechoslovakia 
\def\caa{\refj@jnl{Chinese Astron. Astrophys.}}
                % Chinese Astronomy and Astrophysics
\def\cjaa{\refj@jnl{Chinese J. Astron. Astrophys.}}
                % Chinese Journal of Astronomy and Astrophysics
\def\icarus{\refj@jnl{Icarus}}           % Icarus
\def\jcap{\refj@jnl{J. Cosmology Astropart. Phys.}}
                % Journal of Cosmology and Astroparticle Physics
\def\jrasc{\refj@jnl{JRASC}}             % Journal of the RAS of Canada
\def\memras{\refj@jnl{MmRAS}}            % Memoirs of the RAS
\def\mnras{\refj@jnl{MNRAS}}             % Monthly Notices of the RAS
\def\na{\refj@jnl{New A}}                % New Astronomy
\def\nar{\refj@jnl{New A Rev.}}          % New Astronomy Review
\def\pra{\refj@jnl{Phys.~Rev.~A}}        % Physical Review A: General Physics
\def\prb{\refj@jnl{Phys.~Rev.~B}}        % Physical Review B: Solid State
\def\prc{\refj@jnl{Phys.~Rev.~C}}        % Physical Review C
\def\prd{\refj@jnl{Phys.~Rev.~D}}        % Physical Review D
\def\pre{\refj@jnl{Phys.~Rev.~E}}        % Physical Review E
\def\prl{\refj@jnl{Phys.~Rev.~Lett.}}    % Physical Review Letters
\def\pasa{\refj@jnl{PASA}}               % Publications of the Astron. Soc. of Australia
\def\pasp{\refj@jnl{PASP}}               % Publications of the ASP
\def\pasj{\refj@jnl{PASJ}}               % Publications of the ASJ
\def\rmxaa{\refj@jnl{Rev. Mexicana Astron. Astrofis.}}%
                % Revista Mexicana de Astronomia y Astrofisica
\def\qjras{\refj@jnl{QJRAS}}             % Quarterly Journal of the RAS
\def\skytel{\refj@jnl{S\&T}}             % Sky and Telescope
\def\solphys{\refj@jnl{Sol.~Phys.}}      % Solar Physics
\def\sovast{\refj@jnl{Soviet~Ast.}}      % Soviet Astronomy
\def\ssr{\refj@jnl{Space~Sci.~Rev.}}     % Space Science Reviews
\def\zap{\refj@jnl{ZAp}}                 % Zeitschrift fuer Astrophysik
\def\nat{\refj@jnl{Nature}}              % Nature
\def\iaucirc{\refj@jnl{IAU~Circ.}}       % IAU Cirulars
\def\aplett{\refj@jnl{Astrophys.~Lett.}} % Astrophysics Letters
\def\apspr{\refj@jnl{Astrophys.~Space~Phys.~Res.}}
                % Astrophysics Space Physics Research
\def\bain{\refj@jnl{Bull.~Astron.~Inst.~Netherlands}} 
                % Bulletin Astronomical Institute of the Netherlands
\def\fcp{\refj@jnl{Fund.~Cosmic~Phys.}}  % Fundamental Cosmic Physics
\def\gca{\refj@jnl{Geochim.~Cosmochim.~Acta}}   % Geochimica Cosmochimica Acta
\def\grl{\refj@jnl{Geophys.~Res.~Lett.}} % Geophysics Research Letters
\def\jcp{\refj@jnl{J.~Chem.~Phys.}}      % Journal of Chemical Physics
\def\jgr{\refj@jnl{J.~Geophys.~Res.}}    % Journal of Geophysics Research
\def\jqsrt{\refj@jnl{J.~Quant.~Spec.~Radiat.~Transf.}}
                % Journal of Quantitiative Spectroscopy and Radiative Transfer
\def\memsai{\refj@jnl{Mem.~Soc.~Astron.~Italiana}}
                % Mem. Societa Astronomica Italiana
\def\nphysa{\refj@jnl{Nucl.~Phys.~A}}   % Nuclear Physics A
\def\physrep{\refj@jnl{Phys.~Rep.}}   % Physics Reports
\def\physscr{\refj@jnl{Phys.~Scr}}   % Physica Scripta
\def\planss{\refj@jnl{Planet.~Space~Sci.}}   % Planetary Space Science
\def\procspie{\refj@jnl{Proc.~SPIE}}   % Proceedings of the SPIE

\let\astap=\aap
\let\apjlett=\apjl
\let\apjsupp=\apjs
\let\applopt=\ao

%----------------------------------------------------------------------------------------
%	TITLE PAGE
%----------------------------------------------------------------------------------------
\thispagestyle{empty}
{\Huge Large Synoptic Survey Telescope} 
\linebreak 
\linebreak 
{\Huge Galaxies Science Roadmap}
\linebreak
\linebreak
{\centering
Robertson, Brant E.$^{1}$, 
Banerji, Manda$^{2}$,
Cooper, Michael C.$^{3}$,
Davies, Roger$^{4}$,
Driver, Simon P.$^{5}$,
Ferguson, Annette M. N.$^{6}$,
Ferguson, Henry C.$^{7}$,
Gawiser, Eric$^{8}$,
Kaviraj, Sugata$^{9}$,
Knapen, Johan H.$^{10,11}$,
Lintott, Chris$^{4}$,
Lotz, Jennifer$^{7}$,
Newman, Jeffrey A.$^{12}$,
Norman, Dara J.$^{13}$,
Padilla, Nelson$^{14}$,
Schmidt, Samuel J.$^{15}$,
Smith, Graham P.,$^{16}$,
Tyson, J. Anthony$^{15}$,
Verma, Aprajita$^{4}$,
Zehavi, Idit$^{17}$,
Armus, Lee$^{18}$,
Avestruz, Camille$^{19}$,
Barrientos, L. Felipe$^{14}$,
Bowler, Rebecca A. A.$^{4}$,
Bremer, Malcolm N.$^{20}$,
Conselice, Christopher J.$^{21}$,
Davies, Jonathan$^{22}$,
Demarco, Ricardo$^{23}$,
Dickinson, Mark E.$^{13}$,
Galaz, Gaspar$^{14}$,
Grazian, Andrea$^{24}$,
Holwerda, Benne W.$^{25}$,
Jarvis, Matt J.$^{4,26}$,
Kasliwal, Vishal$^{27,28,29}$,
Lacerna, Ivan$^{30,14}$,
Loveday, Jon$^{31}$,
Marshall, Phil$^{32}$,
Merlin, Emiliano$^{24}$,
Napolitano, Nicola R.$^{33}$,
Puzia, Thomas H.$^{14}$,
Robotham, Aaron$^{5}$,
Salim, Samir$^{34}$,
Sereno, Mauro$^{35}$,
Snyder, Gregory F.$^{7}$,
Stott, John P.$^{36}$,
Tissera, Patricia B.$^{37}$,
Werner, Norbert$^{38,39,40}$,
Yoachim, Peter$^{41}$,
Borne, Kirk D.$^{42}$,
and Members of the LSST Galaxies Science Collaboration 

%\vspace*{2mm}

{\justify\it\small
$^{1}$Department of Astronomy and Astrophysics, University of California, Santa Cruz, Santa Cruz, CA 96054, USA,
$^{2}$Institute of Astronomy, Kavli Institute for Cosmology, University of Cambridge, Madingley Road, Cambridge CB30HA, UK,
$^{3}$Department of Physics and Astronomy, University of California, Irvine, 4129 Frederick Reines Hall, Irvine, CA 92697, USA,
$^{4}$Department of Physics, University of Oxford, Denys Wilkinson Building, Keble Rd., Oxford, OX1 3RH, UK,
$^{5}$International Centre for Radio Astronomy Research (ICRAR), University of Western Australia, Perth, Australia, WA 6009, Australia,
$^{6}$Institute for Astronomy, University of Edinburgh, Royal Observatory, Blackford Hill, Edinburgh, EH9 3HJ, UK,
$^{7}$Space Telescope Science Institute, 3700 San Martin Drive, Baltimore MD 21218, USA,
$^{8}$Rutgers University, 136 Frelinghuysen Rd., Piscataway, NJ 08854-8019, USA,
$^{9}$Centre for Astrophysics Research, University of Hertfordshire, College Lane, Hatfield, Herts AL10 9AB, UK,
$^{10}$Instituto de Astrof\'\i sica de Canarias, E-38200 La Laguna, Spain,
$^{11}$Departamento de Astrof\'\i sica, Universidad de La Laguna, E-38206 La Laguna, Spain,
$^{12}$Department of Physics and Astronomy and PITT PACC, University of Pittsburgh, 3941 O{'}Hara St., Pittsburgh, PA 15260, USA,
$^{13}$NOAO, 950 N. Cherry Ave, Tucson, AZ 85719, USA,
$^{14}$ Instituto de Astrof\'\i sica, Pontificia Universidad,
Cat{\'{o}}lica Chile, Vicu{\~{n}}a Mackenna 4860, Santiago, Chile,
$^{15}$Department of Physics, University of California, Davis, One Shields Ave, Davis, CA, 95616, USA,
$^{16}$School of Physics and Astronomy, University of Birmingham, Edgbaston, B15 2TT, UK,
$^{17}$Department of Astronomy, Case Western Reserve University, 10900 Euclid Avenue, Cleveland, OH 44106, USA,
$^{18}$IPAC/Caltech, 1200 E. California Blvd. MS314-6, Pasadena, CA 91125, USA,
$^{19}$Kavli Institute for Cosmological Physics, University of Chicago, 5640 South Ellis Ave., Chicago, IL 60637, USA,
$^{20}$H.H. Wills Physics Laboratory, University of Bristol, Tyndall Avenue, Bristol, BS8 1TL, UK,
$^{21}$School of Physics and Astronomy, University of Nottingham, Nottingham, NG7 2RD, UK,
$^{22}$Cardiff University, School of Physics and Astronomy, The Parade, Cardiff, CF22 3AA, UK,
$^{23}$Departamento de Astronom\'\i a, Universidad de Concepci{\'{o}}n,Casilla 160-C, Concepci{\'{o}}n, Chile,
$^{24}$INAF - Osservatorio Astronomico di Roma, Via Frascati, 33, I-00078, Monte Porzio Catone (Roma), Italy,
$^{25}$Department of Physics and Astronomy, 102 Natural Science Building, University of Louisville, Louisville KY 40292, USA,
$^{26}$Department of Physics, University of the Western Cape, Bellville 7535, South Africa,
$^{27}$Colfax International, 750 Palomar Avenue, Sunnyvale, CA 94085, USA,
$^{28}$University of Pennsylvania, Department of Physics \& Astronomy, 209 S 33rd St, Philadelphia, PA 19104, USA,
$^{29}$Princeton University, Department of Astrophysical Sciences, 4 Ivy Lane, Princeton, NJ 08544, USA,
$^{30}$Instituto Milenio de Astrof\'\i sica, Av. Vicu{\~{n}}a Mackenna 4860, Macul, Santiago, Chile,
$^{31}$Astronomy Centre, University of Sussex, Falmer, Brighton, BN1 9QH, UK,
$^{32}$Kavli Institute for Particle Astrophysics and Cosmology, P.O. Box 20450, MS29, Stanford, CA 94309, USA,
$^{33}$INAF -Osservatorio Astronomico di Capodimonte, Salita Moiariello, 16, 80131 Naples, Italy, 
$^{34}$Indiana University, Department of Astronomy, Bloomington, IN 47405, USA,
$^{35}$INAF - Osservatorio Astronomico di Bologna; Dipartimento di Fisica e Astronomia, Universit\`a di Bologna Alma-Mater, via Piero Gobetti 93/3, I-40129 Bologna, Italy,
$^{36}$Department of Physics, Lancaster University, Lancaster LA1 4YB, UK,
$^{37}$Astrophysics Group, Department of Physics - Campus La Casona, Universidad Andres Bello, Fernandez Concha 700, Las Condes,  Santiago, Chile,
$^{38}$1 MTA-Eotvos University Hot Universe Research Group, Pazmany Peter Setany 1/A, Budapest, 1117, Hungary,
$^{39}$Department of Theoretical Physics and Astrophysics, Faculty of Science, Masaryk University, Kotlarska 2, Brno, 611 37, Czech Republic,
$^{40}$School of Science, Hiroshima University, 1-3-1 Kagamiyama, Higashi-Hiroshima 739-8526, Japan,
$^{41}$University of Washington, Box 351580, U.W. Seattle, WA 98195-1580, USA,
$^{42}$Booz Allen Hamilton, 308 Sentinel Drive, Suite 100, Annapolis Junction, MD 20701, USA

}

}
\vfill

%----------------------------------------------------------------------------------------
%	COPYRIGHT PAGE
%----------------------------------------------------------------------------------------
\newpage
\thispagestyle{empty}

\noindent
{\justify
The {\it Large Synoptic Survey Telescope Galaxies Science Roadmap} represents the collective efforts of more than one hundred scientists to define the critical research activities to prepare our field to maximize
the science return of the LSST dataset. We want to thank the LSST Corporation for their
support in developing this Roadmap and for supporting LSST-related science more broadly.
We also wish to thank the LSST Galaxies Science Collaboration members for their efforts
over the years in developing the case for extragalactic science with LSST. 
}
\vspace{1in}

Inquiries about this report or its content can be addressed to Brant Robertson ({\tt brant@ucsc.edu}) and the LSST Galaxies Science Collaboration ({\tt lsst-galaxies@lsstcorp.org}).
\vspace{1in}
\begin{center}
Version 1.0:  
August 4, 2017
\end{center}

%----------------------------------------------------------------------------------------
% Abstract
%----------------------------------------------------------------------------------------

\vspace*{30mm}
\begin{center}
{\bf Abstract} 
\end{center}
\vspace*{5mm}

{\justify
The Large Synoptic Survey Telescope (LSST) will enable revolutionary studies of
galaxies, dark matter, and black holes over cosmic time. The
LSST Galaxies Science Collaboration (LSST GSC) has identified a host of preparatory research tasks required 
to leverage fully the LSST dataset for extragalactic science beyond the study of dark energy.
This {\it Galaxies Science Roadmap} provides a brief introduction to critical extragalactic science to be conducted ahead of LSST operations, and a detailed list of preparatory science tasks including the motivation, activities, and deliverables associated with each. The {\it Galaxies Science Roadmap} will serve as a guiding document for researchers interested in conducting extragalactic science in anticipation of the forthcoming LSST era.
}

%----------------------------------------------------------------------------------------
% Table of Contents
%----------------------------------------------------------------------------------------

\tableofcontents % Print the table of contents 

%----------------------------------------------------------------------------------------
%	Chapters
%----------------------------------------------------------------------------------------

% LSST Extragalactic Roadmap
% Chapter: Introduction
% First draft by 

\makeatletter
\let\savedchap\@makechapterhead
\def\@makechapterhead{\vspace*{-3cm}\savedchap}
\chapter[Introduction]{Introduction}
\label{ch:intro}
\let\@makechapterhead\savedchap
\makeatletter

{\justify
The Large Synoptic Survey Telescope (LSST) is a wide-field, ground-based
observatory designed to image a substantial fraction of the sky in six optical
bands every few nights. 
The observatory will operate for at least a decade, allowing 
stacked images to detect galaxies to redshifts well beyond unity. LSST and
its Wide Fast Deep and Deep Drilling Fields surveys will meet 
the requirements of a broad range of science goals in astronomy, astrophysics and cosmology
\citep{ivezic2008a}. 
The LSST ranked first among large ground-based initiatives in the
2010 National Academy of Sciences decadal survey in astronomy and astrophysics \citep{nrc2010a},
and will begin survey operations early in the next decade.
This document, the {\it LSST Galaxies Science Roadmap}, outlines critical preparatory research efforts needed 
to leverage fully the power of LSST for extragalactic science. 

In 2008, eleven separate quasi-independent science collaborations were formed to
focus on a broad range of topics in astronomy and cosmology that the LSST could
address. Members of these collaborations have proven instrumental in helping to
develop the science case for LSST (encapsulated in the LSST Science Book;
\citealt{LSSTSciBook}), 
refine the concepts for the survey and for the data processing, and educate
other scientists and the public about the promise of this unique observatory.

The Dark Energy Science Collaboration (DESC) has taken the
next logical step beyond the Science Book. They identified they most critical
challenges the community will need to overcome
to realize the potential of LSST for
measuring the nature and effects of dark energy. The
DESC looked at five complementary
techniques for tackling dark energy, and outlined high-priority tasks for their
Science Collaboration during construction. The DESC designated sixteen 
new and existing working
groups to coordinate the work. The DESC documented these efforts
in a 133-page white paper \citep{LSSTDESC}. The DESC white
paper provides a guide for investigators looking for ways to contribute to the
overall DESC preparatory science effort, 
indicates clearly the importance of the advance work, and 
connects individual research projects together into a broader 
enterprise to enable dark energy science with LSST.

Following the lead of DESC several other LSST community organizations, 
including the AGN, Milky Way and Local Volume, and Solar System Science
Collaborations,
started to develop Roadmaps to outline critical preparatory tasks in
their science domains.
This document, led by members of the LSST Galaxies Science Collaboration, 
acts as a Roadmap for extragalactic science covering
galaxy formation and evolution writ large, the influence of dark matter structure
formation on the properties of galaxy populations, and the impact of supermassive
black holes on their host galaxies.
This Roadmap identifies the major high-level
science themes of these investigations, outlines how complementary techniques
will contribute, and identifies areas where advance work will prove essential. For this
preparatory work, the {\it LSST Galaxies Science Roadmap} emphasizes areas that are not adequately 
covered in the DESC Roadmap. 

Chapter \ref{ch:science_background} gives a brief summary of the LSST galaxies science background.
Many of the themes and projects are already set out in the LSST Science Book, 
which provides more details for many of the science investigations. 
Chapter \ref{ch:task_lists} presents preparatory science tasks for 
extragalactic science with LSST, organized by science topic.
Cross-references between complementary tasks in different science topics are
noted throughout the document.
The science task list content assumes that the work plan of the DESC will be executed
and that the resulting software and other data products resulting from the DESC
efforts will be made available to the other science collaborations.
}
\let\cleardoublepage\clearpage

% LSST Galaxies Science Roadmap
% Chapter: science_background

\makeatletter
\let\savedchap\@makechapterhead
\def\@makechapterhead{\vspace*{-3cm}\savedchap}
\chapter[Galaxy Evolution Studies with LSST]{Galaxy Evolution Studies with LSST}
\label{ch:science_background}
\let\@makechapterhead\savedchap
\makeatletter

{\justify

Galaxies comprise one of the most fundamental classes of astronomical objects. 
The large luminosities of galaxies enable their 
detection to extreme distances, providing abundant
and far-reaching probes into the depths of the universe.
At each epoch in cosmological history, the color
and brightness distributions of the galaxy population
reveal how stellar populations form with time and
as a function of galaxy mass. The progressive mix of
disk and spheroidal morphological components of 
galaxies communicate the relative importance of
energy dissipation and collisionless processes
for their formation.
Correlations between internal galaxy properties and
cosmic environments indicate
the ways the universe nurtures galaxies as they form.
The evolution of the
detailed characteristics of galaxies over cosmic time
reflects how fundamental astrophysics
operates to generate the rich variety of 
astronomical structures observed today.

Study of the astrophysics of galaxy formation represents
a vital science of its own, but the ready
observability of galaxies critically enables a host of
astronomical experiments in other fields. 
Galaxies act as the semaphores of the
universe, encoding information about
the development of large scale
structures and the mass-energy budget of the
universe in their spatial distribution. The mass distribution
and clustering of galaxies reflect essential
properties of dark matter, including potential
constraints on the velocity and mass of particle candidates.
Galaxies famously host supermassive black holes, 
and observations of active galactic nuclei provide
a window into the high-energy astrophysics of black hole
accretion processes. The porous interface between the
astrophysics of black holes, galaxies, and 
dark matter structures allows for astronomers to 
achieve gains in each field using the same datasets.

LSST will provide a 
digital image of the southern sky in six bands ($ugrizy$).
The area ($\sim18,000~\mathrm{deg}^2$) and depth 
($r\sim24.5$ for a single visit, $r\sim27.5$ coadded) of
the survey will enable research of such breadth
that LSST may influence essentially all extragalactic 
science programs that rely primarily on photometric data.
For studies of galaxies, LSST will provide both an unequaled 
catalogue of billions of extragalactic sources and high-quality 
multiband imaging of individual objects. This section of
the {\it LSST Galaxies Science Roadmap} presents scientific
background for studies of these galaxies with LSST to provide a
context for considering how the astronomical community can
best leverage the catalogue and imaging datasets and for
identifying required preparatory science tasks.

LSST will begin science operations during the next decade,
more than twenty years after the start of the Sloan
Digital Sky Survey \citep{york2000a} and subsequent precursor surveys
including PanSTARRS \citep{kaiser2010a}, the Subaru
survey with Hyper Suprime-Cam \citep{miyazaki2012a}, the
Kilo-Degree Survey \citep{dejong2015a}, and the Dark
Energy Survey \citep{flaugher2005a}. Relative to these prior
efforts, extragalactic science breakthroughs
generated by LSST will likely benefit from its increased area, source
counts and statistical samples, the constraining power of the
six-band imaging, and the survey depth and image quality. The following
discussion of LSST efforts focusing on the astrophysics of galaxies
will highlight how these features of the survey enable new science
programs.

\section{Star Formation and Stellar Populations in Galaxies}
\label{sec:sci:gal:bkgnd:stars}

Light emitted by stellar populations will
provide all the direct measurements made by
LSST. This information will be filtered through
the six passbands utilized by the survey,
providing constraints on the
rest-frame ultraviolet SEDs of galaxies to
redshift $z\sim6$ and a probe of rest-frame
optical spectral breaks to $z\sim1.5$. By
using stellar population synthesis modeling,
these measures of galaxy SEDS will enable 
estimates of the redshifts, star formation rates,
stellar masses, dust content, and 
population ages for potentially 
billions of galaxies. In the context of previous
extragalactic surveys, LSST
will enable new advances in our understanding
of stellar populations in galaxies by contributing
previously unachievable statistical power and an
areal coverage that samples the rarest cosmic
environments.

A variety of ground- and space-based observations
have constrained the
star formation history of the universe over the
redshift range that LSST will likely probe
\citep[for a recent review, see][]{madau2014a}.
The statistical power of LSST will improve our
knowledge of the evolving UV luminosity function,
luminosity density, and cosmic
star formation rate. The LSST observations can
constrain how the astrophysics of gas
cooling within dark matter halos, the efficiency
of molecular cloud formation and the star formation
within them, and
regulatory mechanisms like supernova and radiative
heating give rise to these statistical features
of the galaxy population. While measurement of
the evolving UV luminosity function can
help quantify the role of these 
astrophysical processes, the ability of LSST
to probe vastly different cosmic environments
will also allow for the robust quantification of any
changes in the UV luminosity function with
environmental density, and an examination of 
connections between environment and the fueling
of star formation.

Optical observations teach us about
the established stellar content of galaxies.  
For stellar populations older than $\sim100$ million
years, optical observations provide 
sensitivity to the spectral breaks near a
wavelength of $\lambda\approx4000\mbox{\normalfont\AA}$ in the 
rest-frame related to absorption in the
atmospheres of mature stars. 
Such observations help constrain
the amount of stellar mass in galaxies. For
passive galaxies that lack vigorous star formation,
these optical observations reveal
the well-defined ``red sequence'' of
galaxies in the color-magnitude plane
that traces the succession of
galaxies from recently-merged spheroids
to the most massive systems at the
centers of galaxy clusters \citep[e.g.][]{kaviraj2005a}. For blue,
star-forming
galaxies, optical light can help
quantify the relative contribution of
evolved stars to total galaxy luminosity, 
and indeed has
led to the identification of a well-defined
locus of galaxies in the parameter space of
star formation rate and stellar mass 
\citep[e.g.,][]{noeske2007a}. This
relation, often called the ``star-forming
main sequence'' of galaxies, indicates that
galaxies of the same stellar mass typically
sustain a similar star-formation rate. 
Determining the
physical or possibly statistical 
origin of the relation remains an active
line of inquiry \citep[e.g.][]{lofthouse2017a}, guided by recently improved
data from Hubble Space Telescope over the
$\sim0.2$ deg$^{2}$ Cosmic Assembly Near-Infrared
Deep Extragalactic Survey 
\citep{grogin2011a,koekemoer2011a}. While 
LSST will be comparably limited in redshift 
selection, its $\sim30,000$ times larger area
will enable a much fuller sampling of the
star formation--stellar mass plane, allowing
for a characterization of the distribution
of galaxies that lie off the main sequence
that can help discriminate between phenomenological
explanations of the sequence.

\section{Galaxies as Cosmic Structures}
\label{sec:sci:gal:bkgnd:structures}

The structural properties of galaxies arise from
an intricate combination of important astrophysical
processes. Driven by dark matter structure growth, the dynamical
interplay between baryonic and dark matter components form
the basis for the development of galaxy properties.
The gaseous disks of galaxies require
substantial energy dissipation while depositing
angular momentum into a rotating structure. These
gaseous disks form stars with a
surface density that declines exponentially with
galactic radius, populating stellar orbits that
differentially rotate about the galactic center and
somehow organize into spiral features.
Many disk galaxies contain (pseudo-)bulges that form through
a combination of violent relaxation and orbital dynamics.
These disk galaxy features contrast with systems where
spheroidal stellar distributions dominate the galactic
structure. Massive ellipticals form through galaxy
mergers and accretions, and manage to forge a regular
sequence of surface density, size, and stellar velocity
dispersion from the chaos of strong gravitational
encounters. Since these astrophysical
processes may operate with great
variety as a function of galaxy mass and
cosmic environment, LSST will revolutionize studies
of evolving galaxy morphologies by providing enormous
samples with deep imaging of exquisite quality. These data also enable 
studies of galaxy mass profiles via weak lensing of the background
galaxy population.

The huge sample of galaxies provided by LSST will
provide a definitive view of how the sizes and
structural parameters of disk and spheroidal systems
vary with color, total mass, stellar mass, and luminosity. 
Morphological studies will employ several complementary techniques for quantifying the 
structural properties of galaxies. Bayesian
methods can yield multi-component
parameterized models for all the galaxies
in the LSST sample, including the quantified
contribution of bulge, disk, and
spheroid structures to the observed galaxy
surface brightness profiles. The parameterized
models will supplement non-parametric measures
of the light distribution including the
Gini and M20 metrics that quantify the surface
brightness uniformity and spatial moment of
dominant pixels in a galaxy image \citep{abraham2003a,lotz2004a}. Given the volume 
and steadily increasing depth of the LSST dataset, 
new machine-learning algorithms \citep[e.g.][; Hausen \& Robertson, in prep]{hocking2015a} that enable fast morphological classifications of the LSST survey will be critical in enabling morphological studies from this unique dataset. Collectively, these morphological measures provided
by analyzing the LSST imaging data will enable
a consummate determination of the relation between
structural properties and other features of
galaxies over a range of galaxy mass and luminosity
previously unattainable. 

While the size of the LSST sample supplies the
statistical power for definitive morphological studies,
the sample size also enables the identification of rare
objects. This capability will benefit our efforts for
connecting the distribution of galaxy morphologies to their
evolutionary origin during the structure formation process,
including the formation of disk galaxies.
The emergence of ordered disk galaxies remains a hallmark
event in cosmic history, with so-called ``grand design''
spirals like the Milky Way forming dynamically cold, thin
disks in the last $\sim10$ Gyr. Before thin disks emerged,
rotating systems featured ``clumpy'' mass distributions with
density enhancements
that may originate from large scale gravitational instability.
Whether the ground-based LSST can effectively probe
the exact timing and duration of the transition from
clumpy to well-ordered disks remains 
unknown, but LSST can undoubtedly contribute to studying the
variation in forming disk structures at the present day.
Unusual objects, such as the UV luminous local galaxies identified
by \citet{heckman2005a} that display physical features analogous to 
Lyman break galaxies at higher redshifts, may provide a 
means to study the formation of disks in the present day
under rare conditions only well-probed by the sheer size
of the LSST survey.

Similarly, characterizing the extremes of the
massive spheroid population can critically inform
theoretical models for their formation. For instance,
the most massive galaxies at the centers of galaxy clusters
contain vast numbers of stars within enormous stellar
envelopes. The definitive LSST sample can capture enough
of the most massive, rare clusters to quantify the 
spatial extent of these galaxies at
low surface brightnesses, where the bound stellar
structures blend with the intracluster light of
their hosts. 
LSST data
can improve understanding of
the central densities of local
ellipticals that have seemingly decreased compared with
field ellipticals at higher redshifts. The transformation
of these dense, early ellipticals to the spheroids in the
present day may involve galaxy mergers and environmental
effects, two astrophysical processes that LSST can characterize
through unparalleled statistics and environmental probes.
By measuring the
surface brightness profiles of billions of 
ellipticals LSST can determine whether any such dense
early ellipticals survive to the present day, whatever
their rarity.

Beyond the statistical advances enabled by LSST and the
wide variation in environments probed by a survey
of half the sky, the image quality of LSST will permit
studies of galaxy structures in the very low surface
brightness regime. This capability
will allow for the characterization of stellar halos that
surround nearby galaxies. Structures in stellar halos,
such as tidal features produced by mergers and interactions and density inhomogeneities, originate
from the hierarchical formation process and their
morphological properties provides critical clues to the formation history
on a galaxy-by-galaxy basis \citep{bullock2005a,johnston2008a}.
Observational studies using small, deep surveys like the SDSS Stripe 82 \citep[e.g.][]{kaviraj2014a,kaviraj2014b} and recent work using small telescopes \citep{martinez-delgado2008a,atkinson2013,abraham2014a,van_dokkum2014a}
have 
demonstrated the critical importance of probing the low surface brightness universe 
in order to test the hierarchical galaxy formation paradigm. 
Since low-mass galaxies far outnumber their massive counterparts, the assembly history of massive galaxies is dominated by mergers of unequal mass ratios (`minor' mergers). 
However, such mergers typically produce tidal features that are fainter 
than the surface brightness limits of current surveys like the SDSS. 
Hence, the majority of merging remains, from an empirical point of view, unquantified. 
Deep-wide surveys like LSST are crucial for empirically testing the hierarchical paradigm 
and understanding the role of galaxy merging in driving star formation, black hole growth, and morphological transformations over cosmic time \citep{kaviraj2014b}. 
The examination of stellar halos around galaxies will
result in the identification of small mass satellites
whose sizes, luminosities, and abundances can constrain
the nature of dark matter and models of the galaxy formation process at the extreme
low-mass end of the mass function.  

Finally, observational measures of the outermost regions of thin disks can reveal
how such disks ``end'', how dynamical effects might truncate 
disks, and whether some disks smoothly transition into stellar
halos. LSST will provide such measures and  help quantify the
relative importance of the physical effects that influence the
low surface brightness regions in disks. Other galaxies
have low surface brightnesses throughout their stellar 
structures, and the image quality and sensitivity 
of LSST will enable the most complete census
of low surface brightness galaxies to date. LSST will provide
the best available constraints on the extremes of disk
surface brightness, which relates to the extremes of
star formation in low surface density environments.

The LSST survey uniquely enables precision statistical studies of 
galaxy mass
distribution via weak gravitational.
From the radial dependence of
the galaxy-mass correlation function, galaxy morphological properties 
can be compared with the mass distribution, as a function
of redshift of the lens galaxy population (Choi et al. 2012,
Leauthaud et al. 2012).  Even dwarf galaxies 
can be studied in this way: the LSST survey will enable mass
mapping of samples of hundreds of thousands of dwarf galaxies.
With a sample of hundreds of millions of foreground galaxies, 
for the first time trends in galaxy stellar evolution and type 
can be correlated with halo mass and mass environment on cosmological scales.

\section{Probing the Extremes of Galaxy Formation}
\label{sec:sci:gal:bkgnd:rare}

The deep, multiband imaging that LSST will provide over an enormous
area will enable the search for galaxies that form in the
rarest environments, under the most unusual conditions,
and at very early times. By probing the extremes of
galaxy formation, the LSST data will push our 
understanding of the structure formation process.

The rarest, most massive early galaxies may form in 
conjunction with the supermassive black holes that
power distant quasars. LSST can use the same
types of color-color selections to identify extremely
luminosity galaxies out to redshift $z\sim6$, and
monitor whether the stellar mass build-up in these
galaxies tracks the accretion history of the most
massive supermassive black holes. If stellar mass
builds proportionally to black hole mass in quasars,
then very rare luminous star-forming galaxies at
early times may immediately proceed the formation
of bright quasars. LSST has all the requisite
survey properties (area, multiband imaging, and
depth) to investigate this long-standing problem \citep{robertson2007a}.

The creation of LSST Deep Drilling fields will
enable a precise measurement of the
high-redshift galaxy luminosity function.
Independent determinations of the distribution of 
galaxy luminosities at $z\sim6$ show substantial
variations. The origin of
the discrepancies between various determinations remains
unclear, but the substantial cosmic variance expected
for the limited volumes probed and the intrinsic
rarity of the bright objects may conspire to
introduce large potential differences between
the abundance of massive galaxies in different
areas of the sky. Reducing this uncertainty requires
deep imaging over a wide area, and the LSST Deep Drilling
fields satisfy this need by achieving sensitivities
beyond the rest of the survey. 

The spatial rarity of extreme objects discovered
in the wide LSST area may reflect an intrinsically
small volumetric density of objects or the short duration
of an event that gives rise to the observed properties of the
rare objects. Mergers represent a critical class 
of short-lived epochs in the formation histories of
individual galaxies. Current determinations of the evolving numbers
of close galaxy pairs or morphological indicators of
mergers provide varying estimates for the
redshift dependence of the galaxy merger rate 
\citep[e.g.,][]{darg2010a,conselice2003a,kartaltepe2007a,lotz2008a,lin2008a,kaviraj2009a,robotham2014a,kaviraj2015a}.
The identification of merging
galaxy pairs as a function of separation, merger
mass ratio, and environment in the LSST data will enable
a full accounting of how galaxy mergers influence
the observed properties of galaxies as a function of
cosmic time. 

\vspace{-0.05in}

\section{Photometric Redshifts}
\label{sec:sci:gal:bkgnd:photoz}
As a purely photometric survey, LSST provides an exquisite data set of two-dimensional images of the sky in six passbands.  
However, the third dimension of cosmic distance to each galaxy must often
come from photometric redshifts (photo-$z$'s). Many quantities that LSST data can
reveal about distant galaxy populations, including intrinsic luminosities, physical
sizes, star formation rates, and stellar masses, will ultimately rely on photo-$z$ 
determinations. The engineering of accurate and precise photo-$z$ methods
therefore represents an important science effort for LSST science collaborations.

Spectroscopic distance estimates rely on the identification of atomic or molecular transitions in expensive,
high resolution spectra.  In contrast, photometric redshifts estimate the rough distance to an object based on its broad-band photometric colors and, potentially, other properties measurable from imaging.  
Photo-$z$ measurements are akin to determining redshifts from a very low-resolution but high signal-to-noise spectrum, where each broad-band filter contributes a single sample in that spectrum. Photometric
redshifts are therefore sensitive to the large-scale features of a galaxy spectral energy distribution (e.~g.,~the 4000~\AA\ and Lyman breaks), but in general lack the definitiveness of a redshift measured from multiple well-centroided spectral features (e.g., a pair of emission or absorption lines of known separation). As a result, photometric redshifts will generally be more uncertain than  spectroscopic redshift estimates and can be affected by degeneracies in the color-redshift relation.

By relying on imaging data alone, we will be able to measure photo-$z$'s for billions of galaxies in the LSST survey. 
As errors in the assigned redshift propagate directly to physical quantities of interest, understanding the uncertainties and systematic errors in photo-$z$'s is of the utmost importance for LSST and other photometric surveys. 
Assigning an incorrect redshift to a galaxy also assigns an incorrect luminosity 
owing to misestimation of both the distance modulus and $k$-corrections, and hence can bias estimates of the luminosity function. Errors in redshift will also bias the inferred rest-frame colors of a galaxy, propagating to errors in the inferred spectral type, stellar mass, star formation rate, and other quantities.  
Ideally, estimates of any physical quantity should be performed jointly with a redshift fit, and the expected uncertainties and degeneracies should be fully understood and propagated if measurements are to be
made in an unbiased way.  

To develop optimal estimates of photo-$z$'s  for a particular survey, photo-$z$ algorithms 
should be trained using a set of galaxies with known redshifts. If spectroscopy is obtained for a fully representative sub-sample of the underlying galaxy population spanning the full domain of application, this spectroscopy can also be used to characterize the biases and uncertainties in the photometric redshift estimates, calibrating their use for science.

Obtaining such a fair spectroscopic sample for LSST will be very difficult to achieve due to limitations in instrumentation, telescope time, and the astrophysical properties of galaxies (e.g., weak spectral features). 
Biases owing to incomplete training data can be identified and removed using a variety of redshift calibration techniques, such as spatially cross-correlating photo-$z$-selected datasets with a sample of objects with secure redshifts over wide fields, as will be provided by DESI and 4MOST \citep{newman2008a}.  A detailed plan describing the spectroscopic needs for training and calibrating photometric redshifts for LSST is laid out in \citet[]{newman2015a}, where potential scenarios for obtaining the necessary spectroscopy using existing facilities and those expected to be available in the near future are detailed.   

The insights about the formation and evolution of galaxies we expect to gain from LSST can also be used to improve photo-$z$ algorithms, both by constraining the family of spectral energy distributions of galaxies as a function of redshift and by improving our knowledge of distributions of other observable quantities such as size and surface brightness.  This mutual synergy between understanding galaxy evolution and improved photometric redshift performance should lead to improvements in both areas as the survey progresses.

\vspace{-0.05in}
\section{Science Book}
\label{sec:sci:gal:bkgnd:scibook}

The LSST Science Book (\citealt{LSSTSciBook}) provided 
detailed descriptions of foundational science enabled
by LSST. The LSST Galaxies Science Collaboration authored
the Chapter 9 ``Galaxies'' of the Science Book, and the
table of contents of that chapter follows below to 
provide an example list of topics in extragalactic
science that LSST data will help revolutionize. The
interested reader is referred to the LSST Science 
Book for more details.

\begin{enumerate}
\item Measurements, Detection, Photometry, Morphology
\item Demographics of Galaxy Populations
\begin{itemize}
\item Passively evolving galaxies
\item High-redshift star-forming galaxies
\item Dwarf galaxies
\item Mergers and interactions
\end{itemize}
\item Distribution Functions and Scaling Relations
\begin{itemize}
\item Luminosity and size evolution
\item Relations between observables
\item Quantifying the Biases and Uncertainties
\end{itemize}
\item Galaxies in their Dark-Matter Context
\begin{itemize}
\item Measuring Galaxy Environments with LSST
\item The Galaxy-Halo Connection
\item Clusters and Cluster Galaxy Evolution
\item Probing Galaxy Evolution with Clustering Measurements
\item Measuring Angular Correlations with LSST, Cross-correlations
\end{itemize}
\item Galaxies at Extremely Low Surface Brightness
\begin{itemize}
\item Spiral Galaxies with LSB Disks
\item Dwarf Galaxies
\item Tidal Tails and Streams
\item Intracluster Light
\end{itemize}
\item Wide Area, Multiband Searches for High-Redshift Galaxies
\item Deep Drilling Fields
\item Galaxy Mergers and Merger Rates
\item Special Populations of Galaxies
\item Public Involvement
\end{enumerate}
}

\input{task_lists}

%---------------------------------------------------------------------------------------
% References
%---------------------------------------------------------------------------------------

%---END---
\end{document}